\documentclass[traditabstract,a4]{aa}

\usepackage{graphicx,amssymb,amsmath}
\usepackage{natbib}

\newcommand{\ain}{a_{\rm in}}
\newcommand{\aout}{a_{\rm out}}

\newcommand{\Sigout}{\Sigma_{\rm out}}

\newcommand{\md}{M_{\rm d}}
\newcommand{\psid}{\psi_{\rm d}}

\newcommand{\psiout}{\psi_{\rm out}}

\newcommand{\bindm}{M_{\rm b}}
\newcommand{\tbindm}{\tilde{M}_{\rm b}}
\newcommand{\mstar}{M_{\rm BH}}
\newcommand{\mbh}{M_{\rm BH}}
\newcommand{\tmstar}{\tilde{M}_{\rm BH}}
\newcommand{\tmbh}{\tilde{M}_{\rm BH}}
\newcommand{\tmd}{\tilde{M}_{\rm d}}
\newcommand{\tmu}{\tilde{\mu}}
\newcommand{\rs}{R_{\rm S}}

\begin{document}

\title{AGN disks and black holes on the weighting scales}
\authorrunning{J.-M. Hur\'e et al.}
\titlerunning{AGN disks and black holes on the weighting scales}
\author{Jean-Marc Hur\'e\inst{1,2}, Franck Hersant\inst{1,2}, Cl\'ement Surville\inst{3}, Naomasa Nakai\inst{4}, \and Thierry Jacq\inst{1,2}}

\offprints{jean-marc.hure@obs.u-bordeaux1.fr}

\institute{Universit\'e de Bordeaux, OASU, 2 rue de l'Observatoire, BP 89, F-33271 Floirac Cedex, France
\and
CNRS, UMR 5804, LAB, 2 rue de l'Observatoire, BP 89, F-33271 Floirac Cedex, France
\and
LAM, Universit\'e de Provence, 13388 Marseille, France
\and
Institute of Physics, University of Tsukuba, Ten-noudai, Tsukuba, Ibaraki 305-8571, Japan}
\date{Received ??? / Accepted ???}

\abstract
{We exploit our formula for the gravitational potential of finite size, power-law disks to derive a general expression linking the mass of the black hole in active galactic nuclei (AGN), the mass of
the surrounding disk, its surface density profile (through the power index $s$), and the differential rotation
law. We find that the global rotation curve $v(R)$ of the disk in centrifugal balance does not obey a
power law of the cylindrical radius $R$ (except in the confusing case $s=-2$ that mimics a Keplerian motion),
and discuss the local velocity index. This formula can help to understand how, from
position-velocity diagrams, mass is shared between the disk and the black hole. To this purpose, we have checked
the idea by generating a sample of synthetic data with different levels of Gaussian noise, added in radius. 
It turns out that, when observations are spread over a large radial domain and exhibit low
dispersion (standard deviation $\sigma \lesssim 10\%$ typically), the disk properties (mass and
$s$-parameter) and black hole mass can be deduced from a {\it non linear fit of kinematic data} plotted on
a $(R,R v^2)$-diagram. For $\sigma \gtrsim 10 \%$, masses are estimated fairly well from a {\it linear
regression} (corresponding to the zeroth-order treatment of the formula), but the power index $s$ is no longer
accessible. We have applied the model to $7$ AGN disks whose rotation has already been probed through water maser emission.
For NGC3393 and UGC3789, the masses seem well constrained through the linear approach. For IC1481, the
power-law exponent $s$ can even be deduced. Because the model is scale-free, it applies to any kind of
star/disk system. Extension to disks around young stars showing deviation from Keplerian motion is thus
straightforward.}


\maketitle

\section{Introduction}

The mass of astrophysical objects  | except maybe for stars | is generally difficult to determine with
precision, mostly because of inappropriate tracers, relatively low spatial resolution, and a certain
misunderstanding of the internal structure and physical processes involved. This is the case for giant disks orbiting
supermassive black holes in active galactic nuclei (AGN). For some nearby objects, the cold gas
rotating in the outermost regions (the subparsec scale typically from the center) is detected at radio wavelengths through water vapor emission \citep[e.g.][]{miyoshi95,braatz09}. The inner regions, not accessible yet to current instruments, could host the bulk of the mass if the total
surface density in the disk varies roughly with the cylindrical radius $R$ as $R^{-2}$ or faster \citep{ss73,cd90}. Estimating the
disk mass is a complex task. It necessitates a global disk model capable of describing the dynamics of
gas, its thermodynamics, its chemical complexity, as well as its interaction with radiation (lines and
continuum). The disk mass is an important quantity in
understanding the AGN phenomenon. Along with the accretion rate, turbulent viscosity, and black hole mass, it helps to put
constraints on the activity of the AGN in terms of stability, lifetime, luminosity, and matter supplied from the
host galaxy \cite[][]{combes01,coza08}.

The mass of disks can also be probed via the consequences of their gravity. All the material contained within a disk exerts gravitational forces on itself | the so-called ``self-gravity''| which influences or even strongly governs (like in galaxies) internal orbital motions, sometimes up to instability \cite[e.g.][]{mestel63,binneytremaine87,linpap95}. Even in the presence of a massive central object, self-gravity may cause a slight deviation in Kepler's law, which is interesting to analyze and to quantify. Obviously, non-Keplerian rotation can have other origins like pressure effects such as in slim/thick disks \citep{ab88} or magnetic fields \citep{heyp89}. Here, we focus on self-gravity, which is expected to play a role in geometrically thin disks \citep{sw82,sb87}.

There are many articles that aims to establish the relation between the dynamics (through gravitational potentials) and mass density distribution, especially in the context of galactic dynamics \citep{binneytremaine87}. Existing potential/density pairs do not however seem fully appropriate to gaseous disks surrounding a central object, probably because star/disk systems where the gas exhibit non-Keplerian motions are still marginal. As models and theories suggest \citep[e.g.][]{ss73,pringle81,cd90,hure98}, gaseous disks in AGN are large and are expected to exhibit a self-similar behavior over some radial range. \cite{hureetal08} has determined an accurate formula for the gravitational potential in the midplane of a flat power-law disk with {\it finite size and mass}. It is valid for a wide range of the power index for surface density. In this article, we use this result to derive {\it an algebraic relation} between the orbital velocity of the gas, the disk parameters (surface density profile, mass, size), and the mass of the central object, assuming a pressure-less disk at centrifugal equilibrium, as commonly done. This relation furnishes a simple method for
determining how the mass is shared between the disk and the central object. As expected, the ``modified''
rotation law is not, in this model, a power law of the radius as often considered in this context \citep{herrnstein05}. Although this study is valid for any kind of astrophysical star/disk system (like in circumstellar environments), we focus on AGN disks whose
kinematics have been observed in VLBI through water-maser emission.

This article is organized as follows. In Section \ref{sec:model}, we recall the model of a pressure-less
disk at centrifugal equilibrium surrounding a central black hole. We introduce the formula for the gravitational
potential in the disk midplane by \cite{hureetal08}, and derive the general expression for the velocity of the
orbiting gas as a function of the disk mass, black hole mass, and surface density profile through
the power-law index. In Section \ref{sec:massdetmeth}, we show how
this expression (or its zero-order version) can be used to estimate how the mass is shared between a central object and its surrounding disk by analyzing observational data in a ``position-dynamical mass'' diagram (instead of the classical position-velocity diagram). We first applied the method to IC1481, thereby refining the disk parameters reported in \cite{mamyoda09}. We discuss uncertainties in Sect. \ref{sec:uncertainties}, and show how dispersion naturally goes against the method. Section \ref{sec:application} is devoted to applying of the method to a sample of a few well known AGN hosting a masing outer disk. We conclude in Section \ref{sec:sum}.

\section{The basic model}
\label{sec:model}

\subsection{A pressure-less disk at centrifugal equilibrium}

We consider a gaseous disk with inner edge $\ain$ and outer edge $\aout \gg \ain$, orbiting a central black
hole with mass $\mstar$. This disk is assumed to be axially symmetrical, flat (i.e. no vertical thickness),
pressure-less, and steady. At centrifugal equilibrium, the rotation velocity $v$ of material at cylindrical
distance $R$ in the midplane of the disk, in the reference frame of the black hole, is given by the standard relation:
\begin{equation}
v^2(R)= \frac{G \mstar}{R} + R \frac{d \psid}{d R},
\label{eq:vphi2}
\end{equation}
where $\psid$ is the gravitational potential of the disk. This latter function critically depends  on the surface density profile $\Sigma(R)$ through the Poisson integral. It is generally not easily accessed by analytical means, even in the actual one-dimensional case.

There is a broad literature devoted to determining potential-density pairs $(\psid,\Sigma)$ for axially symmetric systems \citep[e.g.][]{binneytremaine87, evanscollett93}. Here, we consider the class of flat, power-law distributions where the surface density varies according to
\begin{equation}
\label{eq:sigma}
\Sigma=
\begin{cases}
\Sigout \varpi^s \quad \text{if} \, \varpi \in [\Delta,1],\\
0 \quad \text{elsewhere},
\end{cases}
\end{equation}
where $R = \aout \varpi$, $\Delta = \ain/\aout$ is the axis ratio, $\Sigout$ the surface density at the outer edge, and $s$ is a constant. Such a profile seems well-suited for large, gaseous disks in AGN, at least in the framework of geometrically thin disk models that predict $s \approx -1$ typically \citep{ss73,pringle81,cd90,hure98}. Potential-density pairs for flat power-law disks, including Mestel's solution ($s=-1$), are summarized in \cite{evansread98}. These correspond to infinite disks (i.e. $\ain=0$ and $\aout\rightarrow \infty$) whose mass is infinite as soon as $s>-2$. \cite{conway00} has produced formal solutions corresponding to finite disks and no inner edge (i.e. $\ain=0$), but for even positive indexes (i.e. $s=0,2,...$). Unfortunately, accounting for edges increases the mathematical difficulties. \cite{hureetal08} have recently produced a reliable approximation for $\psid$ associated with Eq. \ref{eq:sigma}, namely (see their Eq. 53)
\begin{equation}
\label{eq:psiaxwithax}
- \frac{\psi_d(\varpi)}{\psiout} \approx  B \varpi^{1+s} + \frac{1}{\varpi} \frac{\varpi^{2+s}-\Delta^{2+s}}{2+s} + \frac{1-\varpi^{1+s}}{1+s},
\end{equation}
where\footnote{This constant $\psiout$ has dimension of a potential, but it is not the value at the outer edge.} $\psiout = 2 \pi G \Sigout \aout$ and
\begin{equation}
B = \frac{6C-\pi-1}{\pi}\approx 0.431,
\label{eq:B}
\end{equation}
where $C$ is the Catalan's constant. This approximation is accurate within a few percent, provided the disk is large enough ($\Delta \ll 1$) and $-3~\lesssim~s~\lesssim~0$. These conditions are probably met in most astrophysical disks, especially in AGN disks \citep{ss73,pringle81,cd90,hure98}, and others \citep[e.g.][]{dubrulle92}. Actually, we have $\Delta \sim 3 \rs/\aout  \approx 10^{-5}$ for a $10^8 M_\odot$ AGN black hole accreting a parsec size disk ($\rs$ being the Schwarzschild radius). Besides, Eq. \ref{eq:psiaxwithax} does not suffer from the edge singularities expected when considering sharp edges.

\begin{figure}
\includegraphics[width=9.cm, , bb=18 43 741 584, clip=]{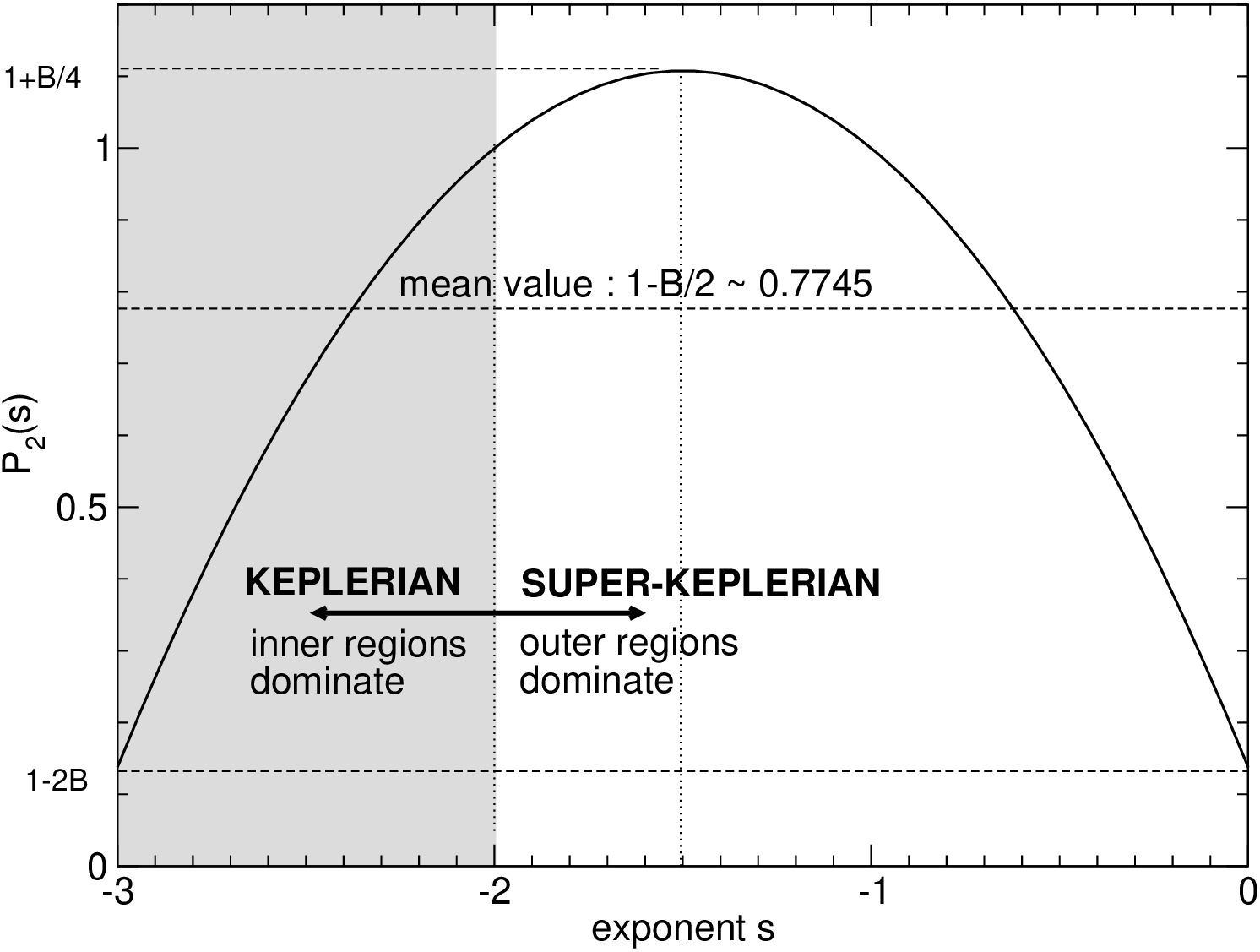} 
\caption{$P_2(s)$ in the range of validity of Eq. \ref{eq:psiaxwithax}.}
\label{fig:pofs.eps}
\end{figure}

\subsection{The dynamical mass}

As Eq. \ref{eq:vphi2} shows, a ``good'' variable to measure masses in this model is \citep[e.g.][]{yamauchi04}:
\begin{equation}
\mu=\frac{R v^2}{G} \equiv \mu(\varpi),
\label{eq:dynmass}
\end{equation}
and the disk makes its own contribution through $\psid$. In the case of a spherical distribution (where $R \rightarrow r$), this dynamical mass would represent the enclosed mass at a given radius \citep{herrnstein05}. Here, things are quite different since matter is gathered in a plane. There is no obvious use of the Gauss theorem, although the monopole approximation gives the right order of magnitude. Inserting in Eq. \ref{eq:dynmass} the velocity $v$ given by Eq. \ref{eq:vphi2}, and using Eq. \ref{eq:psiaxwithax} for the disk potential $\psid$, we finally get
\begin{equation}
\label{eq:o}
\mu(\varpi) = \mstar - \md \frac{ \Delta^{2+s} - \varpi^{s+2} P_2(s)}{1 - \Delta^{2+s}},
\end{equation}
where $\md$ is the disk mass (at $\aout$), and 
\begin{equation}
P_2(s)=1- B(1+s)(2+s)
\end{equation}
is a second-order polynomial in $s$. It is displayed in Fig. \ref{fig:pofs.eps}. In the range of interest,
$P_2(s)$ is always positive and has a limited range of variation, since $0.14 \lesssim P_2(s) \lesssim 1.11$.
We also have $P_2(-2)=P_2(-1)=1$ and a mean value of $\sim 0.8$.

From Eq. \ref{eq:o}, we expect two extreme behaviors of the function $\mu(\varpi)$ depending on the disk properties. If $s \le -2$ (the case of "centrally peaked" distributions), we have
\begin{equation}
\mu \sim \mstar + \md = cst.
\label{eq:o1}
\end{equation}
In the absence of radial gradient of $\mu$, it is not possible to separate the disk and the central black
hole, {\it regardless of the disk mass}. A Keplerian rotation curve result from a massive disk (without central
object) or to a massive central object (with a light disk, as often considered). If $s>-2$ (the bulk of the disk mass stands in the outermost regions), we have

\begin{equation}
\mu(\varpi) \sim \mstar + \md  \varpi^{s+2} P_2(s),
\label{eq:o2}
\end{equation}
which is essentially an increasing function of the radius (positive gradient). Disk rotation is therefore always {\it super-Keplerian}. The larger the disk mass, the larger the deviation from Kepler's law.

We conclude that, if the disk mass is significant with respect to the central mass, $v^2(R)$ is the sum of two
power laws of the radius, and {\it this sum is not a power law} (see below). In other words, fitting the global rotation
curve of a system containing a massive disk and a black hole with a single power law (e.g. $v\propto R^\gamma$), cannot give any quantitative information about the mass distribution in the framework of Newtonian gravity. This approach is often considered when a massive disk is suspected
\citep[see e.g.][]{herrnstein05,kondratko08,mccallum09}. See, however, appendix \ref{app} for a short discussion of the velocity index $\gamma$.

 For $s=-1$, Eq. \ref{eq:o} reads
\begin{equation}
\label{eq:osmoins1}
\mu(\varpi) = \mstar + \md \frac{\varpi - \Delta}{1 - \Delta},
\end{equation}
which is to be compared to the case of a Mestel disk \citep{mestel63}
\begin{equation}
\label{eq:mestel}
\mu_{\rm Mestel}(\varpi) = \md'(\varpi),
\end{equation}
where $\md'$ is the cumulative disk mass at the actual radius (linear with the radius $\varpi$). As already pointed out elsewhere \citep{binneytremaine87}, this equation ``happens to give the same answer'' as what is deduced from the Gauss theorem for a spherically symmetric distribution. In contrast, Eq. \ref{eq:osmoins1} corresponds to a Mestel disk truncated on both sides. It includes {\it edge effects and total disk} mass, and  {\it explicitly contains the central point mass} (which is not part of Mestel's disk model). In the limit $\Delta \rightarrow 0$, we then recover Mestel's solution.

\section{AGN disk/black hole mass determination method. The case of IC1481}
\label{sec:massdetmeth}

\subsection{Position-dynamical mass diagram}

We immediately see from above that, if the rotation curve of the disk is partly known in the form of $N$ observational points $\{(R_i,v_i)\}$, then some constraints can be set on the disk mass, black hole mass, and surface density profile by fitting the data $\{(R_i,\mu_i)\}$ through Eqs. \ref{eq:o1} or \ref{eq:o2}. Obviously, this procedure does not guarantee that the triplet $(\mstar,\md,s)$ is physically meaningful given the simplicity of our model and assumptions. Uncertainties in data also fragilize the inversion. Thus, there are three different possibilities.
\begin{enumerate}
\item[A:] Data points $\{(R_i,\mu_i)\}_N$ show {\it no noticeable variation} around a constant value, only a certain dispersion. The systems thus appears in Keplerian rotation. We deduce that either there is a light disk surrounding a massive black hole or the disk is rather massive but the gas is distributed such that $s\le-2$. The diagram only gives the quantity $\mstar+\md$, which can be identified with the so-called ``binding mass'' $\bindm$ (or enclosed mass). There is no way to separate the black hole and the disk in this analysis.
\item[B:] Data points show {\it a significant variation}, still with a certain dispersion (see Sect. \ref{sec:uncertainties}). The gas rotates faster than Keplerian. If $\mu$ increases faster than $\varpi$, then $s > -1$, otherwise $s < -1$. In either case, fitting the data points through Eq. \ref{eq:o2} can yield a triplet ($\mstar$, $\md$, $s$).
\item[C:] Data cannot be fitted by Eq. \ref{eq:o2}, or inferred parameters are non physical. In this case, our model is inappropriate. Various reasons can be invoked (see Sect. \ref{sec:sum}). 
\end{enumerate}

\subsection{Zeroth-order: disk mass and central mass}

The zeroth-order treatment of the non linear formula is interesting and instructive because it gives the
orders of magnitude. Actually, if we consider that astrophysical disks are characterized by $s \approx -1$, we can expand Eq. \ref{eq:o2} around $s=-1$. We find (see also Eq. \ref{eq:osmoins1} with $\Delta \rightarrow 0$)
\begin{flalign}
\nonumber
\mu & \approx \mstar + \md \varpi + \md(s+1) \varpi \ln \varpi,\\
 & \approx \mbh + \md \varpi.
\label{eq:mus}
\end{flalign}
We conclude that, if observational data plotted on a position-dynamical mass diagram are almost linearly distributed in a position-dynamical mass diagram, then {\it the slope is the disk mass} $\md$ and {\it the intercept} is the black hole mass. In the following, we discuss both approaches in the context of AGN disks whose rotation, for some of them, is known from maser emission.

\begin{figure}
\includegraphics[width=8.cm, bb=7 0 596 787,clip=]{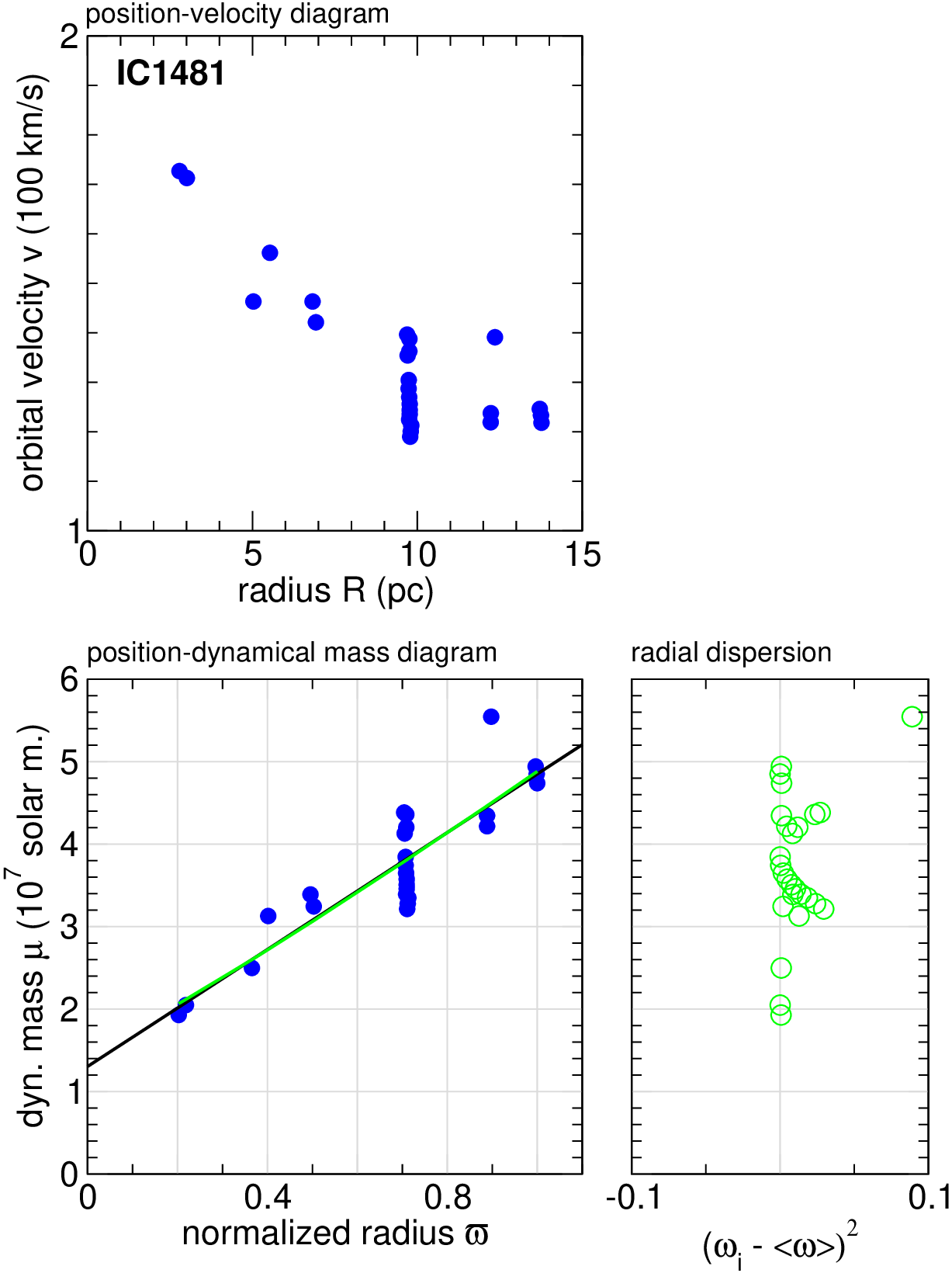} 
\caption{Position-velocity data of maser spots in the disk of IC1481 ({\it top}), and corresponding
position-dynamical mass diagram ({\it bottom}). A non linear fit of the data ({\it green curve}) gives the disk mass, the black hole mass and the $s$-parameter. The linear fit ({\it black line}) is also shown. Data are from \cite{mamyoda09}.}
\label{fig:ic1481} 
\end{figure}

\subsection{Scale-free formula. Scaling to AGN disks}

Equation \ref{eq:o2} is totally scale-free if we divide $\mu$ by the black hole mass. Actually, if $q=\md/\mbh$ denotes the mass ratio in the system, Eq. \ref{eq:o2} writes
\begin{equation}
\frac{\mu(\varpi)}{\mbh} \sim 1 + q  \varpi^{s+2} P_2(s),
\label{eq:o2-scalefree}
\end{equation}
so that the model applies to any kind of star/disk system. In the context of AGN, black holes are supermassive (several million solar masses), the central accretion disk typically has the parsec size, and rotational velocities are, at such distances, hundreds of km/s. With this scaling, the formula for the dynamical mass becomes\footnote{For a circumstellar system, we would have, for instance,
\begin{equation}
\frac{\mu_i}{1 M_\odot} \approx 1.12  \left(\frac{R_i}{100 \, \rm AU}\right) \left( \frac{v_i}{100 \, \rm km/s} \right)^2.
\end{equation}}
\begin{flalign}
\frac{\mu_i}{10^7 M_\odot} &\approx 0.232  \left(\frac{R_i}{1 \, \rm pc}\right) \left( \frac{v_i}{100 \, \rm km/s} \right)^2 \\
& \equiv \tilde{\mu}_i
\end{flalign}
where $v_i$ is, following our model, the disk rotational velocity measured relative to the systemic/receding velocity of the system, and $R_i$ relative to the rotation axis of the disk. In the following, masses are expressed in units of $10^7 M_\odot$, and denoted $\tilde{M}$. Then, Eq. \ref{eq:o2} becomes
\begin{equation}
\tilde{\mu}(\varpi) = \tmstar + \tmd  \varpi^{s+2} P_2(s),
\label{eq:to2}
\end{equation}
and its linear version is
\begin{equation}
\tmu(\varpi) = \tmbh + \tmd \varpi.
\label{eq:tmus}
\end{equation}

\subsection{An example: the case of IC1481}

The top panel in Fig. \ref{fig:ic1481} shows the rotational velocity of maser spots observed in the outer
disk of the active nucleus in galaxy IC1481 versus the distance from the center \citep{mamyoda09}.  The radius
of the outermost maser spot is set to $\aout$ (although the gas disk can extend farther away\footnote{It can
be shown that Eqs. \ref{eq:o1} and \ref{eq:o2} still hold if one uses a reference value $a_0 \le \aout$ other
than the outer radius $\aout$ since $\md \propto R^{s+2}$. Then $\md$ refers to the cumulative disk mass up to
this reference radius $a_0$.}). The bottom panel displays the same data once converted into a
$(\varpi,\mu)-$diagram, as well as the result of the fit by Eq. \ref{eq:to2} (non-linear) and by  Eq. \ref{eq:tmus}
(linear). For each point, the square of the radial deviation between the data and the linear fit is shown (see
below). The parameters of the two fits are gathered in Table \ref{tab:agnnonlinear}. These are in good
agreement, especially because the solution of the non linear fit gives an $s$-parameter close to $-1$. In such
a case, the curvature is difficult to detect by eye. As announced in \cite{mamyoda09}\footnote{In
\cite{mamyoda09}, an extreme case without black hole was considered.}, the disk mass is higher than the black
hole mass by a factor of $2-3$. 

\begin{table}
\centering
\begin{tabular}{cccccccc}
AGN &  points & $\langle \tmu_i \rangle$ & $\tmbh$ & $\tmd$ & $s$ & cor.  \\ \hline
IC1481  & $26$ & $3.67$ & $1.51$ & $3.59$ & $-0.88$ & $0.89$ \\
IC1481 & $26$ & $3.67$ & $1.30$ & $3.55$ & ($-1$) & $0.89$ & \\
 \hline \\
\end{tabular}
\caption{Results of non linear fit through Eq. \ref{eq:to2} for IC1481 ({\it top}), and results for the linear fit (a least-square fit) through Eq. \ref{eq:tmus} ({\it bottom}). Last column is the correlation coefficient. Maser data are from \cite{mamyoda09}. 
} 
\label{tab:agnnonlinear}
\end{table}

\section{Uncertainties and data dispersion}
\label{sec:uncertainties}

Position-velocity data deduced from observations generally suffer from dispersion and uncertainties, which can
have different origins: physical (e.g. variability, non uniform dynamics, geometry and deprojection) and instrumental (i.e.
lack of resolution). In particular, locating the position of emitters precisely is a critical point
\cite[e.g.][]{uscanga07}. In order to check whether the method is ``robust'' enough to infer some reliable
information about masses, we generated a sample of $N$ points $\{\varpi_i,\tmu_i\}$, obeying exactly Eq.
\ref{eq:to2} for a given reference triplet $(\tmstar,\tmd,s)^{\rm ref}$. To these synthetic data, we added
uncertainties on the radius (without presuming their origin) with three different levels of ``noise''. We considered a Gaussian noise, with various
standard deviations $\sigma$. The perturbed radii $\varpi_i$ were then all rescaled so that finally
$\varpi_N =1$. Unfortunately, this rescaling procedure introduce a slight bias by tending to overestimate both
the black hole mass mass and the disk mass.

\begin{figure}[h]
\centering
\includegraphics[height=4.5cm, bb=9 46 400 407, clip=]{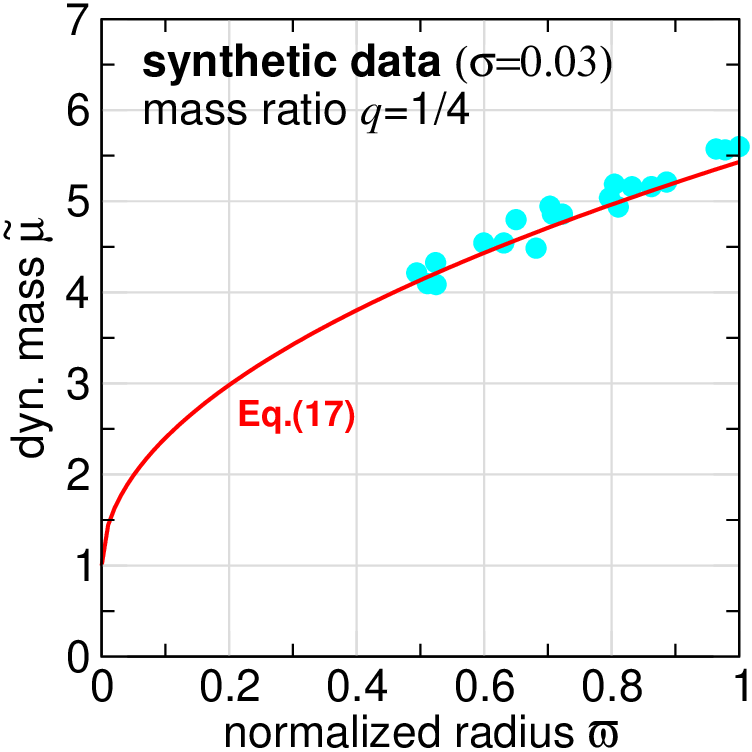} 
\includegraphics[height=4.5cm, bb=9 46 400 407, clip=]{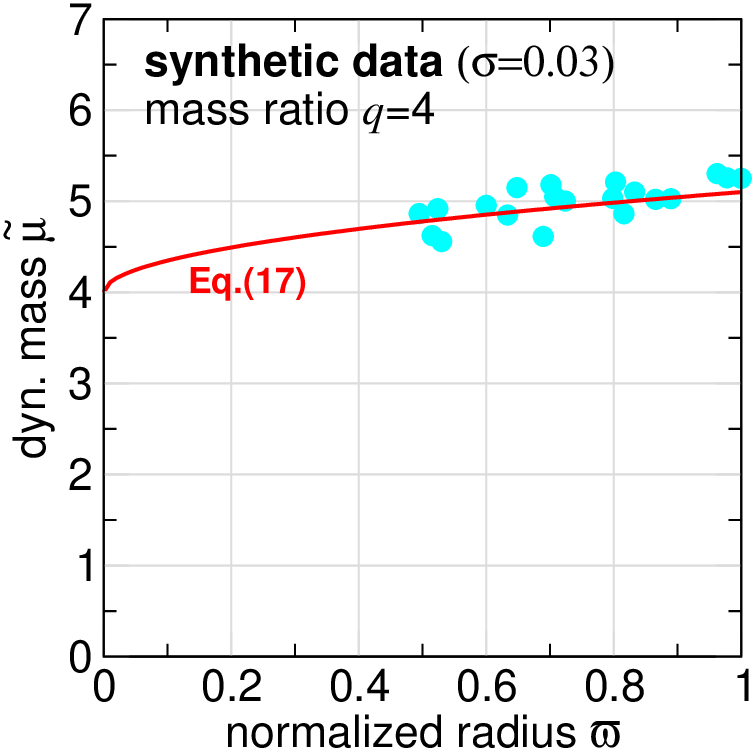} 
\caption{Synthetic data set made with $N=20$ points generated from Eq. \ref{eq:to2} with $(\tmstar,\tmd,s)^{\rm ref} = (1,4,-1.5)$ ({\it top}) and with $(\tmstar,\tmd,s)^{\rm ref} = (4,1,-1.5)$ ({\it bottom}). Uncertainties corresponding to a Gaussian noise with a standard deviation $\sigma=0.03$ have been added to all the radii. 
 Data were rescaled in radius such that the outer point has a normalized radius $\varpi_N=1$.}
\label{fig:syntheticdata}
\end{figure}

\begin{table}
\centering
\begin{tabular}{ccccc}
\multicolumn{5}{l}{Massive disk configuration ($q=4$)}\\ \hline \hline
$\sigma$ & $0.03$ & $0.05$ & $0.1$ & comments \\ \hline
$\delta$ & $1.79$ & $1.67$ & $\mathbf{1.24}$ & linear approach\\
$\tmbh$   & $2.74$ & $2.62$ & $2.19$ &  ($s=-1$)\\
$\tmd$    & $2.89$ & $3.11$ & $3.91$ &  \\\\

$\delta$  &  $\mathbf{0.33}$ & $0.61$ &$1.57$ & non-linear\\
$\tmbh$   & $0.68$ & $0.42$ & $-0.49$ & approach \\
$\tmd$    & $4.43$ & $4.75$ & $5.89$  &  (with $s=-1.5$)\\\\

$\delta$ & $\mathbf{0.32}$  & $0.83$ & no solution &       non-linear\\
$\tmbh$   & $1.31$  & $0.20$ & | & approach\\
$\tmd$    & $3.88$  & $4.95$ & | \\
$s$      & $-1.41$ & $-1.52$ & | \\
 \hline \hline \\\\
\multicolumn{5}{l}{Massive black hole configuration ($q=\frac{1}{4}$)}\\ \hline \hline
$\sigma$ & $0.03$ & $0.05$ & $0.1$ & comments \\ \hline
$\delta$ & $\mathbf{0.34}$ & $0.46$ & $1.44$ & linear approach\\
$\tmbh$   & $4.28$ & $4.06$ & $3.38$ &  ($s=-1$)\\
$\tmd$    & $0.96$ & $1.31$ & $2.39$ &  \\\\

$\delta$  &  $\mathbf{0.49}$ & $1.03$ &$2.70$ & non-linear\\
$\tmbh$   & $3.59$ & $3.13$ & $1.71$ &  approach\\
$\tmd$    & $1.48$ & $2.01$ & $3.64$  & (with $s=-1.5$) \\\\

$\delta$ & $\mathbf{0.20}$  & $4.52$ & no solution &       non-linear\\
$\tmbh$   & $4.07$  & $-0.45$ & | & approach\\
$\tmd$    & $1.09$  & $5.46$ & | \\
$s$      & $-1.24$ & $-1.83$ & | \\
 \hline \hline \\
\end{tabular}
\caption{Results for synthetic data generated from Eq. \ref{eq:to2}, for the massive disk configuration ({\it top}), and for the massive black hole configuration ({\it bottom}). Values in bold correspond to the best approach.} 
\label{tab:synthetic}
\end{table}

Here, we report a typical experiment obtained for
\begin{itemize}
\item a massive disk configuration with $(\tmstar,\tmd)^{\rm ref} = (1,4)$ corresponding to mass ratio $q=4$,
\item a massive black hole configuration with $(\tmstar,\tmd)^{\rm ref} = (4,1)$ corresponding to $q=\frac{1}{4}$, 
\end{itemize}
with $s^{\rm ref}=-1.5$ and $N=20$ points in both cases. Initially, data were randomly spread over the range $[0.5,1]$, and we considered 3 levels of noise: 
$\sigma= \{0.03,0.05,0.1\}$. The synthetic $(\varpi,\tmu)$-diagram obtained for $\sigma=0.03$, which is
presumed to mimic dispersed observationnal data, is shown in Fig. \ref{fig:syntheticdata} with the exact
rotation law given by Eq. \ref{eq:to2}. We tried to determine the best triplet $(\tmstar,\tmd,s)$ in three
ways:

\begin{itemize}
\item by the linear approach,
\item by the non linear approach, but forcing $s=s^{\rm ref}$,
\item by the non linear approach.
\end{itemize}
The second case is only illustrative, as it can not be used in practice since $s$ is not known a priori. For each configuration, the relative distance between the reference triplet and the one deduced by fitting the synthetic noisy data is measured by the parameter $\delta$, with
\begin{equation}
\delta^2 =\left(1- \frac{\tmstar}{\tmstar^{\rm ref}}\right)^2+\left(1- \frac{\tmd}{\tmd^{\rm ref}}\right)^2+\left(1- \frac{s}{s^{\rm ref}}\right)^2.
\end{equation}

As a result, $\delta \rightarrow 0$ when the method works, whereas $\delta \gtrsim 1$ when it fails (i.e. when the
output and the input triplets differ significantly\footnote{This norm can introduce a bias into the
interpretation as, a relative discrepancy of $60\%$ in each direction suffices to produce $\delta=1$, but
$\delta$ can be large even if two of the three parameters are correct.}). Practically, the linear approach is performed from a least-square procedure. The results are gathered in Table
\ref{tab:synthetic}. As expected, the lower the dispersion, the better the method. For the massive disk configuration and $\sigma=0.03$ (low dispersion), we recover the disk mass within a few percent by the non linear fit, while the error in the black hole mass is about $30\%$. By the linear approach, we get the disk mass within $30\%$, while there is a factor $2-3$ for the black hole mass. This is not all that surprising since the curvature of the function $\tmu(\varpi)$ is noticeable in this case, so the error on the intercept can be large. As the dispersion increases, the linear approach becomes more and more reliable for getting the disk mass (within a few percent), but the black hole mass is still poorly determined (a factor $2$ typically). Regarding the massive black hole configuration and $\sigma=0.03$ (low dispersion), the black hole mass and the disk mass are deduced correctly within a few percent typically, especially using the non linear method. As the dispersion increases, the uncertainty in both two quantities rises, for the same reason as mentioned hereabove; however, the black hole mass is determined with about $15\%$.

\begin{table*}
\centering
\begin{tabular}{llccccl}
AGN & Reference & $\tbindm$ & $\tmbh$ & $\tmd$ & $s$ \\ \hline
IC1481 & \cite{mamyoda09} & $-$ & $<1$ & $4.3$ & $-1.38$ & no black hole limit\\
UGC3789 & \cite{2009reid} & $1.1$ & $1.1$ & $-$ &  \\
NGC3393 & \cite{kondratko08} & $3.1$ at $0.36$ pc & $2.6$ & $1.9$ \\
NGC4258 & \cite{herrnstein05} & $3.8$ & $3.8$ & $0.089$& \\
NGC1068  & \cite{1997gg}  & $1.5$ & ? & ? & \\
         & $^\dagger$\cite{lodatobertin03} & $1.5$ & $0.8$ & $0.8$ &  $-1$ &  thin disk solution \\
         & $^\dagger$\cite{hure02} & $-$ & $1.2$ & $0.6$ ($0.8$  at $1.5$ pc) & $-1$ & thin disk solution \\
         &  & $-$ & $1.2$ & $0.9$ ($1.1$ at $1.3$ pc) & $-1.05$ & thick disk solution\\
NGC4945 & \cite{1997gmh} & $0.14$ at $0.3$ pc & ? & $-$  \\
Circinus &\cite{mccallum09} &  $-$ &  $0.17$ & $-$ \\
         & (epoch 1)\\
 \hline \\
\end{tabular}
\caption{List of AGN hosting a masing disk considered here, references for position-velocity data and associated disk/black hole parameters reported by authors ($^\dagger$model only). Columns 3-5 gives values found by the authors. Masses are given in units of $10^7$ M$_\odot$; $\bindm$ is the binding mass as determined from spherically symmetric models.}
\label{tab:agnlist}
\end{table*}

\begin{figure*}

\includegraphics[height=5.cm, bb=20 30 626 400, clip=]{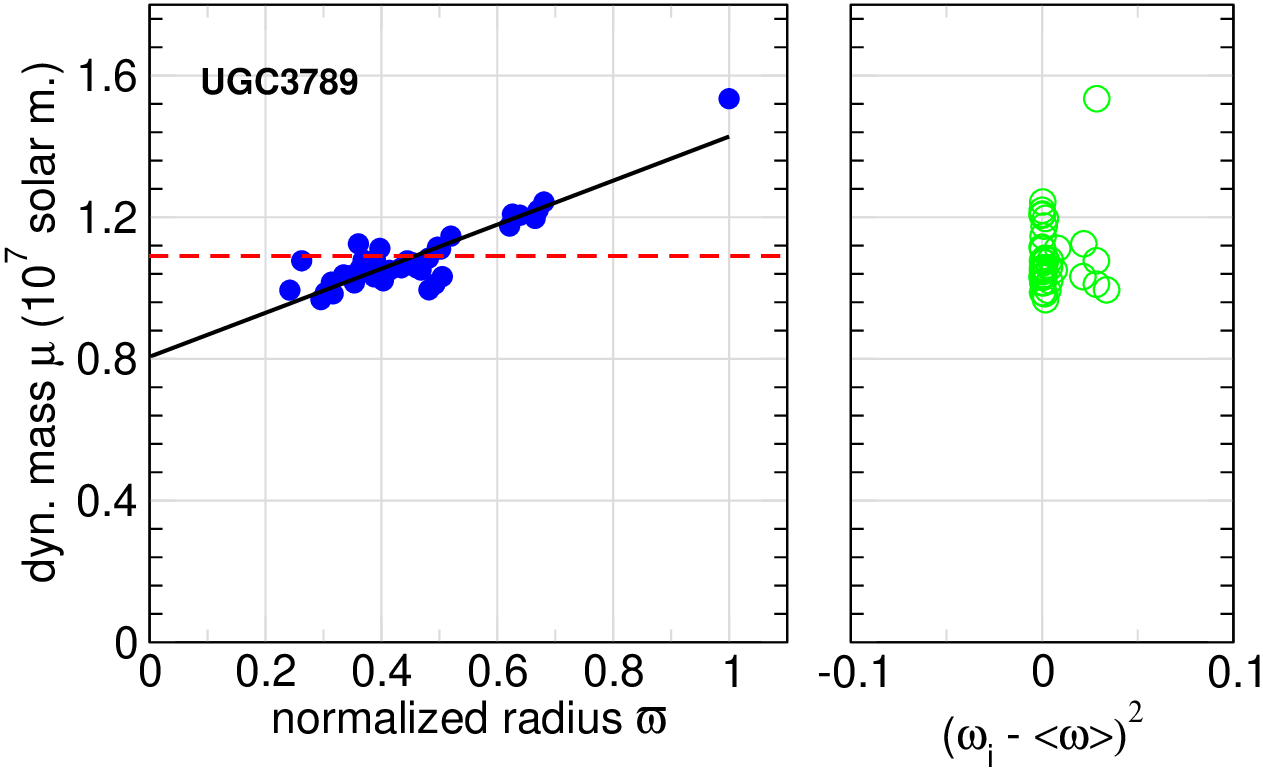} 
\qquad\includegraphics[height=5.cm, bb=9 30 614 407, clip=]{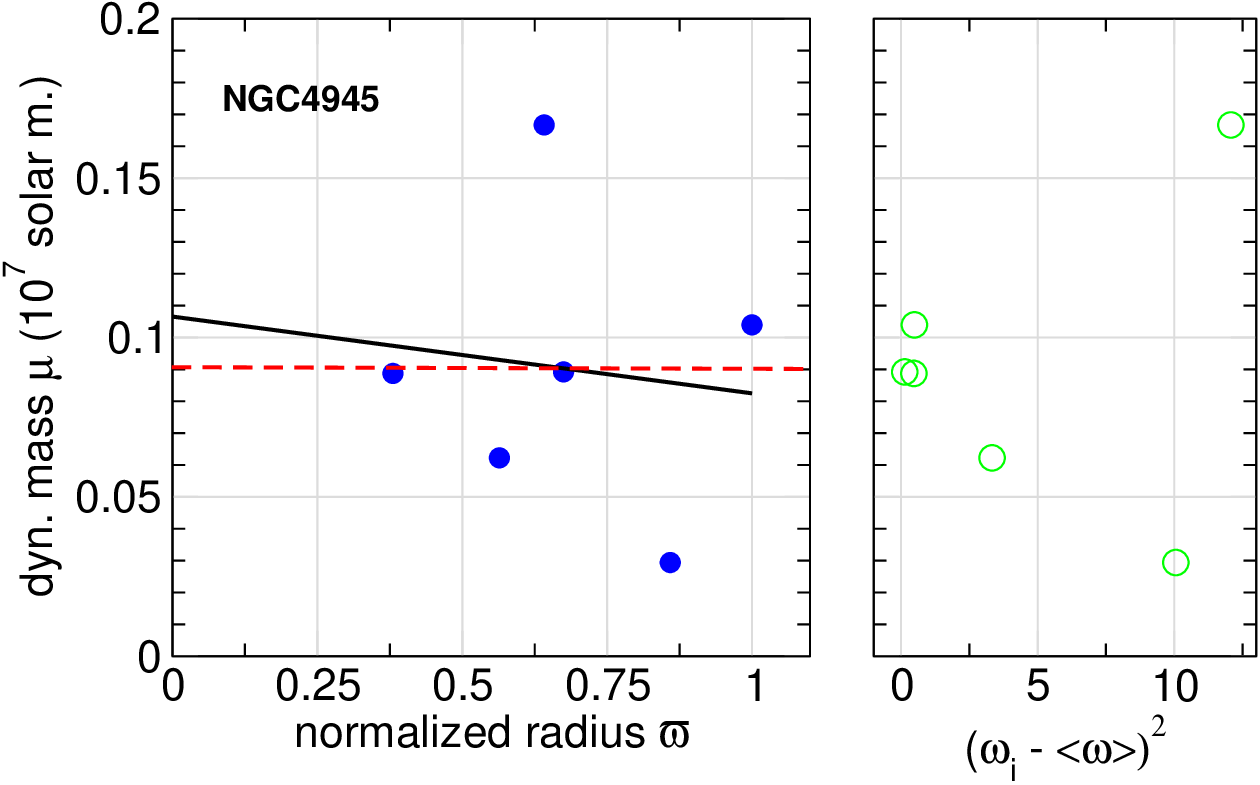} 

\includegraphics[height=5.cm, bb=38 30 634 400, clip=]{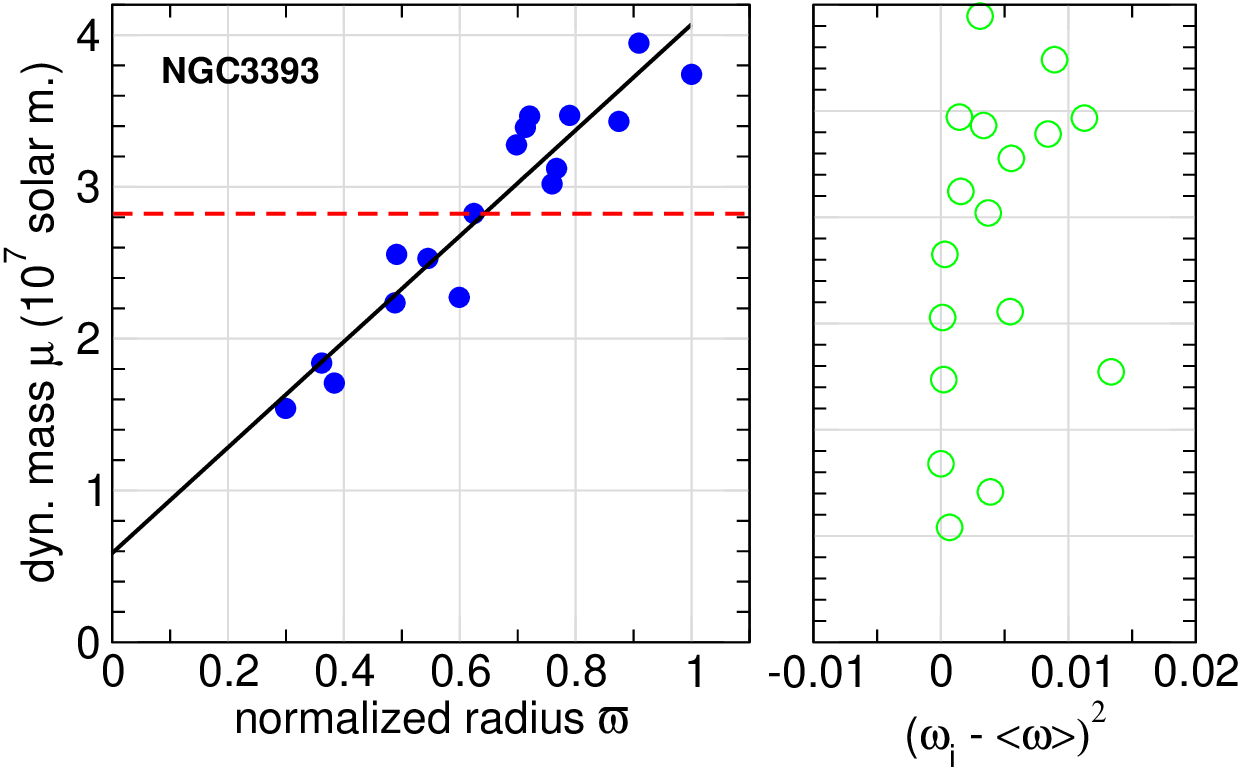} 
\qquad\includegraphics[height=5.cm, bb=20 30 628 400, clip=]{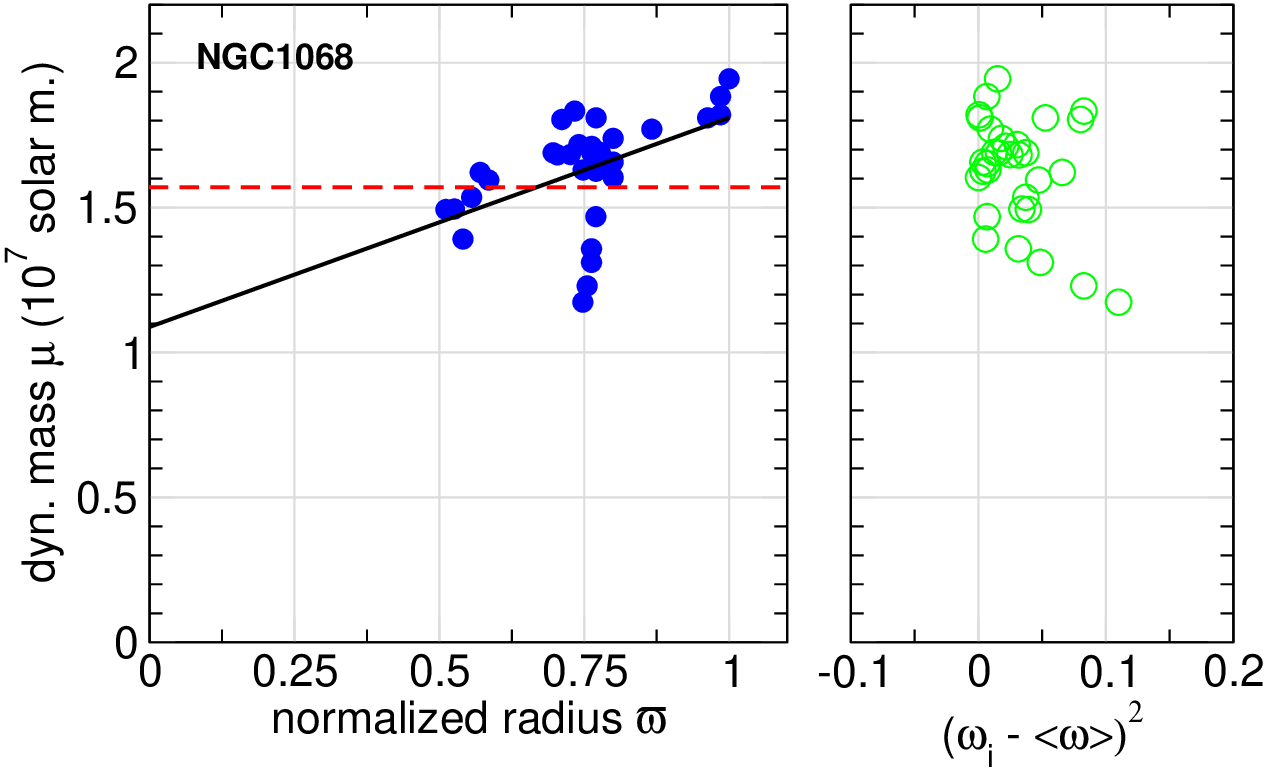} 

\includegraphics[height=5.cm, bb=20 30 628 400, clip=]{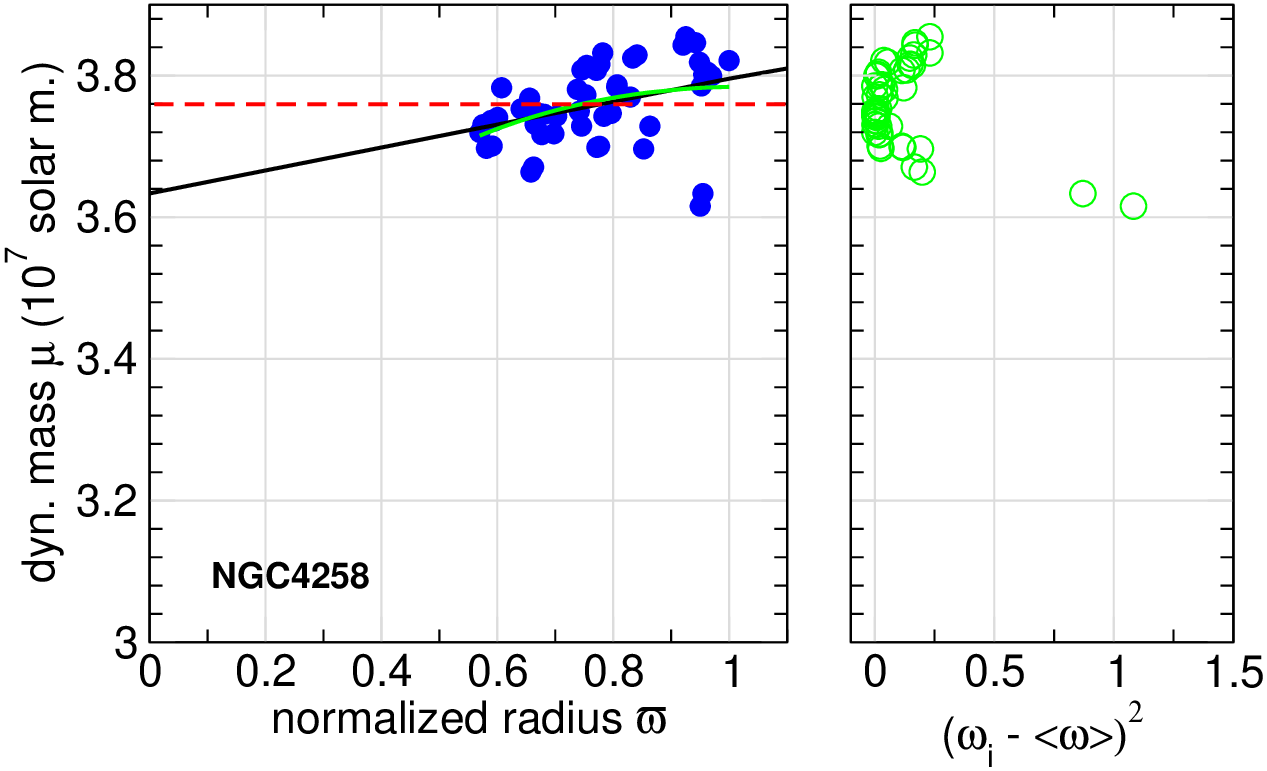} 
\qquad\includegraphics[height=5.cm, bb=20 31 675 400,clip=]{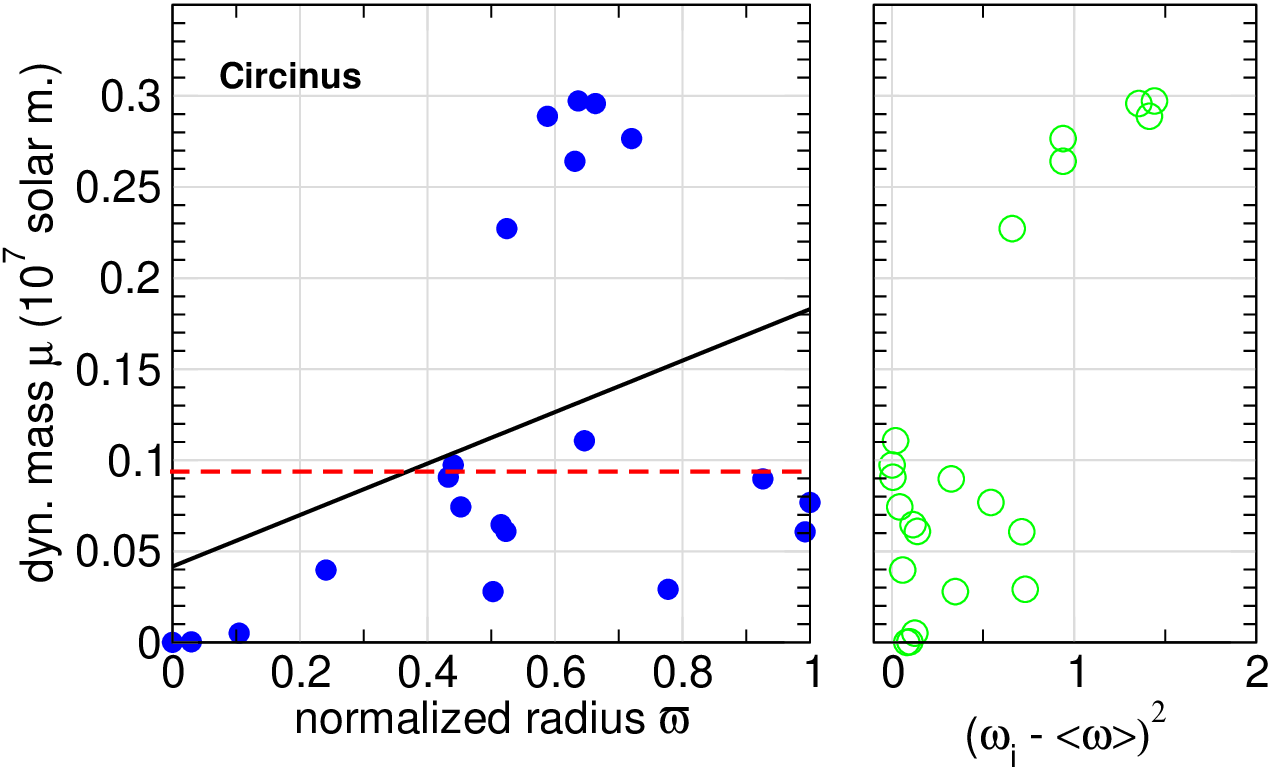} 

\caption{Position-dynamical mass diagram for each AGN listed in Table \ref{tab:agnlist} ({\it left}) and deviation in radius with the linear fitting ({\it right}). Observational data are circles, and the red dashed line is the averaged value of the $\{\tmu_i\}$ considered. The linear fit through Eq. \ref{eq:tmus}  is the bold line (see Table \ref{tab:agnlinear}). }
\label{fig:allagn}
\end{figure*}

\begin{table*}
\centering
\begin{tabular}{ccccccccl}
AGN &  $N$ & $\langle \tmu_i \rangle$ & $\tmbh$ & $\tmd$ & cor. & case$^*$  & st. dev. $\sigma$ & comment for non linear fitting\\ \hline
 IC1481 & $26$ & $3.67$ & $1.30$ & $3.55$ & $0.89$ & B & $0.11$ & agreement\\\\
 UGC3789 & $39$ & $1.09$ & $0.81$ & $0.62$ & $0.89$ & B &  $0.07$ & $(\tmbh,\tmd,s) \approx (5.91,-2.13,-0.056)$\\ 
 NGC3393 & $17$ & $2.85$ & $0.59$ & $3.48$ & $0.95$ & B &  $0.06$ & no solution\\
 NGC4258 & $50$ & $3.78$ & $3.63$ & $0.16$ & $0.37$ & AB & $0.3$ & $(\tmbh,\tmd,s) \approx (3.20,0.60,-1.97)$, cor. $0.38$ \\
 NGC1068 & $32$ & $1.63$ & $0.72$ & $1.09$ & $0.49$ & BC & $0.2$ & $(\tmbh,\tmd,s) \approx (2.77,-0.89,-3.58)$ \\\\
{\it NGC4945} & $6$ & $0.09$ & $0.11$ & $-0.02$  & $-0.11$ & C & $2$  & $(\tmbh,\tmd,s) \approx (0.025,0.053,-3.38)$, cor. $0.14$ $(s < -2)$\\
{\it Circinus} & $21$ & $0.12$ & $0.04$  &$0.14$ & $0.37$ & C & $0.6$ & no solution \\
 \hline \\
\end{tabular}
\caption{Results of linear-fitting through Eq. \ref{eq:tmus}. Column 8 gives the standard deviation $\sigma$. $^*$For the classification, see Sect. \ref{sec:massdetmeth}}
\label{tab:agnlinear}
\end{table*}

 Since the errors on the radius propagate to the variable $\mu$, we made the same fit in the ($R$,$v$)-diagram and found very similar results. In brief, we see that when the ``global slope'' is steep (massive disk configuration), the disk mass is  determined with precision. Conversely, when the ``global slope'' is almost zero  (massive black hole configuration), the black hole mass (value at the intercept) is well determined. When dispersion is too large, the correlation becomes too weak for interpretation, and the method fails to give the triplet, but it is important to note that the linear approach always gives the value of the most massive component within a factor less than about $15 \%$, and the less massive by a factor $2-3$.

\section{Application to a few AGN masing disks}
\label{sec:application}

In the case where position-velocity diagrams are available in the literature, we analyzed a few AGN masing disks to
try to put constraints on masses. This is not a new problem regarding the mass of black hole. The mass of
disks is, however, much a subject of debate, since there is almost no systematic or generic study like the one
presented here, except isolated attempts \citep[see e.g.][]{hure02,lodatobertin03,herrnstein05,kondratko08,mccallum09}. 
Table \ref{tab:agnlist} lists the systems
considered here. Position-velocity data points were obtained by digitalizing graphs when published,
without special treatment (in particular, we fully trust in the analysis by the authors to furnish
deprojected position-velocity data in the reference frame of the black hole). The table also contains masses
previously proposed for the black hole and disk, mostly through models of spherical distributions or models.
In each case listed in Table \ref{tab:agnlist}, we have tried to fit the data by the linear formula and by the
non linear formula as well. Position-dynamical mass diagrams, together with the linear fit and averaged value,
are displayed in Fig. \ref{fig:allagn}. The results are given in Table \ref{tab:agnlinear}. In particular, from the linear regression, we estimated the degree of dispersion by computing the standard deviation $\sigma$ (column $8$). From the simple analysis performed in Sect. \ref{sec:uncertainties}, we immediately expect no precise values for NGC4945 and Circinus. For NGC4258 and NGC1068, the dispersion is much weaker with $\sigma \approx 0.2-0.3$. For these two objects, the linear approach should give the mass of the most massive component  quite correctly (i.e. see Tab. \ref{tab:synthetic}). Finally, for IC1481, UGC3789 and NGC3393, dispersion is less than $\sim 10\%$, so both mass components should be obtained with ``accuracy''.

It was not possible to derive a triplet ($\mstar$, $\md$, $s$) for all these systems through the
non linear approach (negative mass or non convergence of the fitting procedure), even for UGC3789 and NGC3393 whose data,
like for IC1481, are spread over a large radial domain and little dispersed. In two cases (NGC4945 and
Circinus), the method is inappropriate: no correlation really exists in the position-dynamical mass and
inferred parameters are non physical (in particular, $\md <0$ for NGC4945). We are aware that some disks have
been predicted to be geometrically thick, so the model of flat disk used here should naturally fail. In
contrast, results obtained through the linear approach are satisfactory for UGC3786, NGC3393, and NGC1068 where
disk mass is similar to, or higher than the mass of the central black hole. For these objects, the
massive disk should be prone to instabilities. For NGC4258, the slope is very weak and a mass ratio of about
$0.05$ is expected.

\section{Concluding remarks}
\label{sec:sum}

In this paper, we have reported a simple method to estimate the mass of a disk surrounding a central black
hole and assuming a flat, pressure-less disk at centrifugal balance. This model obviously applies to other kinds
of systems containing a disk and (possibly) a central star. Instead of the conventional position-velocity
diagram, the position-dynamical mass diagram makes the measurement of possible deviations to Kepler's law easier.
Using recent calculations for the gravitational potential of ``truncated self-similar'' disks, we have shown
that i) the rotation law is generally not a power law of the radius, and ii) such deviations are, at zero
order at least, directly related to the cumulative disk mass.

It is clear that the disk model is very simple and can be improved in several ways. At the same time, it is
difficult to make the method robust and universal, since position-velocity data generally show a certain
dispersion that can have several origins like thickness effects, instabilities (warps), etc. Through a simple analysis of uncertainties, we have shown that the linear approach gives quite correctly the most massive component (typically with a few tens of percent)  if data dispersion is large (standard deviation larger than $0.1$). For weak dispersion, both the disk mass and the black hole mass should be accessible through non linear data fitting, as well as the surface density power index.
 Moreover,
the systemic velocity  plays a major role, and the asymmetry often observed between the redshifted part and
the blueshifted part of the rotation curve must be accounted for. This method must therefore be seen as a
first step in the analysis of masses in star/disk systems based on gravity.

\begin{acknowledgements}
 We thank S. Collin and F. Herpin for useful comments. The referees are acknowledged for their comments and suggestions, in particular about the problem of uncertainties.
\end{acknowledgements}

\bibliographystyle{aa}
\bibliography{../hurebibtex}

\appendix
\section{Note on the power index of the rotation curve}

\label{app}

Let $\gamma$ be the local index of the rotation velocity, namely
\begin{equation}
v=\sqrt{GM_0}R^\gamma,
\end{equation}
where $M_0$ is a reference mass. In the present model, it is not a constant, but a function of the radius. We easily find from Eq. \ref{eq:dynmass} 
\begin{equation}
\frac{d \ln \mu}{d \ln \varpi} = 1+2\gamma,
\end{equation}
with  $\gamma=-\frac{1}{2}$ for a Keplerian motion associated with a central point mass $M_0$. From Eq. \ref{eq:o2}, we also deduce
\begin{equation}
\frac{d \ln \mu}{d \ln \varpi} = \frac{\md}{\mu}  P_2(s) (s+2) \varpi^{s+2},
\end{equation}
and then $\gamma$ can be written in the form:
\begin{equation}
\gamma = -\frac{1}{2} + \delta \gamma_{\rm kep.},
\end{equation}
where
\begin{equation}
\delta \gamma_{\rm kep.}  = \frac{\md}{2\mu}  P_2(s) (s+2) \varpi^{s+2}
\end{equation}
is the deviation to the Keplerian index. For $s=-2$, we have $ \delta \gamma_{\rm kep.}=0$: the rotation curve resembles a Keplerian curve due to a point mass (see above). This result was already known. The mean deviation is
\begin{equation}
\langle \delta \gamma_{\rm kep.} \rangle = \frac{1}{1-\Delta}\int_\Delta^1{\frac{\md}{2\mu}  P_2(s) (s+2) \varpi^{s+2} d \varpi},
\end{equation}
where $q=\md/\mbh$ is the disk-to-black hole mass ratio. The above expression can be integrated exactly for some values of the $s$-parameter. For instance, with $s=-1$, we find $\langle \delta \gamma_{\rm kep.} \rangle \approx \frac{q}{4}$. It means that a deviation to Kepler's law as low as $0.02$ on the velocity index (i.e. $\gamma = -0.48$) could imply a mass ratio $q=10 \%$. This value is, in magnitude, consistent (within a factor $\sim 2$) with the monopole approximation that predicts a mass ratio $\sim 0.4$. For systems containing a disk significantly less massive than the central object (i.e. $q \lesssim 1$) and $\Delta \ll 1$, we find
\begin{equation}
\langle \delta \gamma_{\rm kep.} \rangle \approx \frac{q}{1+q} \times \frac{(s+2)}{2(s+3)} P_2(s).
\end{equation}

\begin{figure}[h]
\includegraphics[width=9.cm]{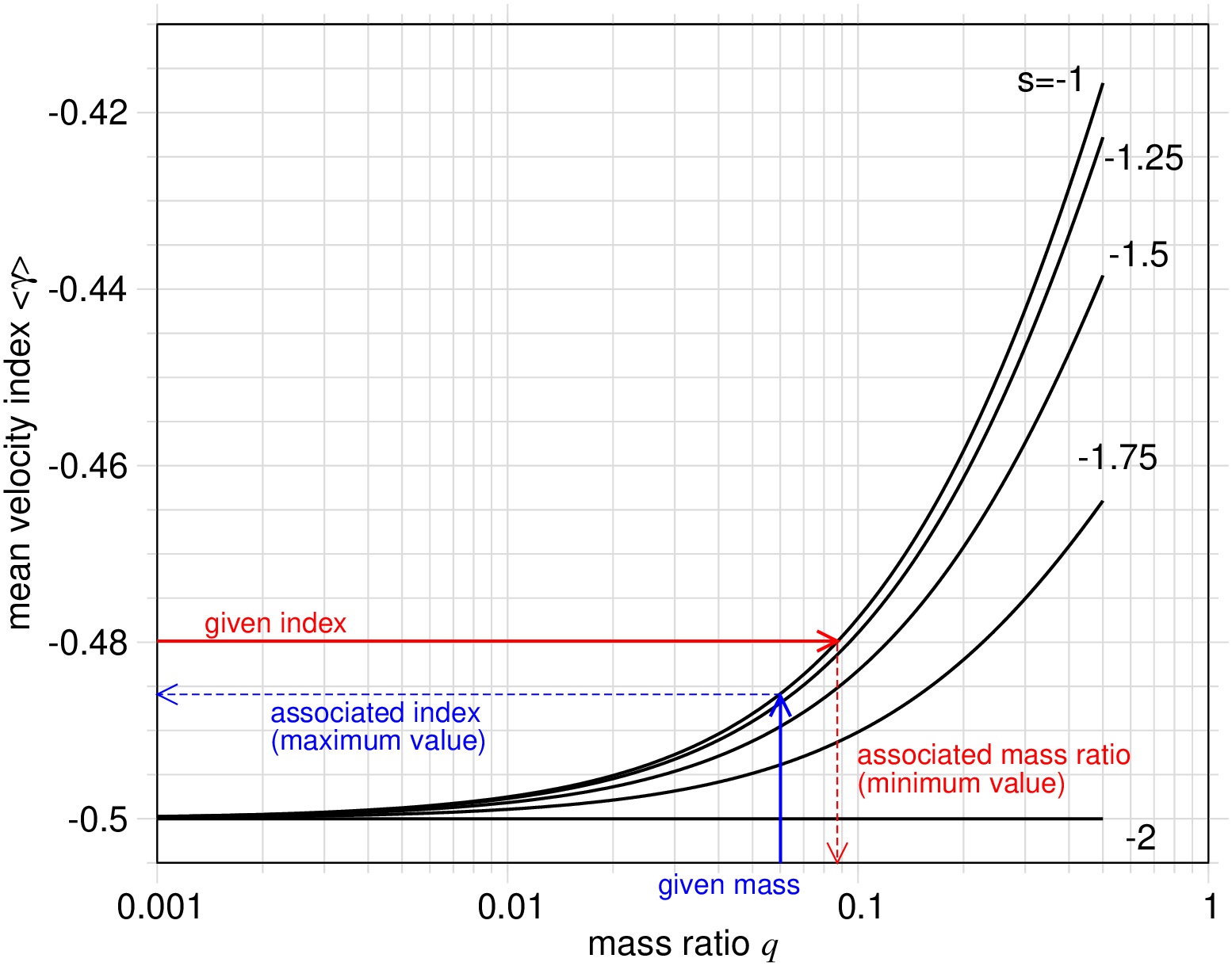} 
\caption{Relation between the mean index of the super-Keplerian rotation law and the disk mass to black hole mass ratio for some values of the $s$-parameter. To a given mean index $\langle \gamma \rangle$, it corresponds a minimum mass ratio $q$ ({\it red}). To a given mass ratio, it corresponds a maximal mean index ({\it blue}).} 
\label{fig:deltakep}
\end{figure}

Figure \ref{fig:deltakep} displays the mean velocity index $\langle \gamma\rangle$ versus $q$ for a few values of the $s$-parameter. This plot enables either to predict the maximum departure to Kepler's law for a given mass or to bound the disk mass given a mean velocity index. Practically, if deviations remain of small amplitude for low-mass disks, these may not be exploitable (data dispersion, thickness effects).

\end{document}